\documentclass[a4paper,fleqn]{cas-dc}

\usepackage[numbers]{natbib}
\usepackage{siunitx}

\usepackage{caption}
\usepackage{lipsum}
\newenvironment{bottompar}{\par\vspace*{\fill}}{\clearpage}

\def\tsc#1{\csdef{#1}{\textsc{\lowercase{#1}}\xspace}}
\tsc{WGM}
\tsc{QE}
\tsc{EP}
\tsc{PMS}
\tsc{BEC}
\tsc{DE}

\begin{document}
\let\WriteBookmarks\relax
\def\floatpagepagefraction{1}
\def\textpagefraction{.001}
\shorttitle{Hysteretic transition in a dipole cluster}
\shortauthors{S.~Völkel, S.~Hartung, I.~Rehberg:}

\title [mode = title]{Comment on "Hysteretic transition between states of a filled hexagonal magnetic dipole cluster"}                      



\author{Simeon Völkel}[orcid=0000-0002-0036-0394]

\author{Stefan Hartung}[orcid=0000-0003-0629-0865]

\author{Ingo Rehberg}[orcid=0000-0001-7511-2910]
\address{Experimentalphysik V, Universit\"at Bayreuth, 95440 Bayreuth, Germany}

\begin{abstract}
In the paper "Andrew D.P.~Smith, Peter T.~Haugen, Boyd F.~Edwards: Hysteretic transition between states of a filled hexagonal magnetic dipole cluster, Journal of Magnetism and Magnetic Materials 549 (2022): 168991" a hysteretic transition between two stable arrangements of a cluster of seven dipoles is presented.
The relative strength of the center dipole in a hexagonal arrangement serves as the bifurcation parameter.
The authors clearly demonstrate the existence of two instabilities accompanied by discontinuous jumps of the dipole arrangement, but leave the question about the nature of these instabilities unanswered.
This comment clarifies the nature of the two instabilities:  the first one is a symmetry-breaking sub-critical bifurcation with parabolic scaling of the magnetic potential energy difference between the two branches, and the second one is a fold with its characteristic scaling in the form of a semi-cubic parabola.
\end{abstract}

\begin{graphicalabstract}
\begin{figure}
\includegraphics [width=0.95\linewidth]{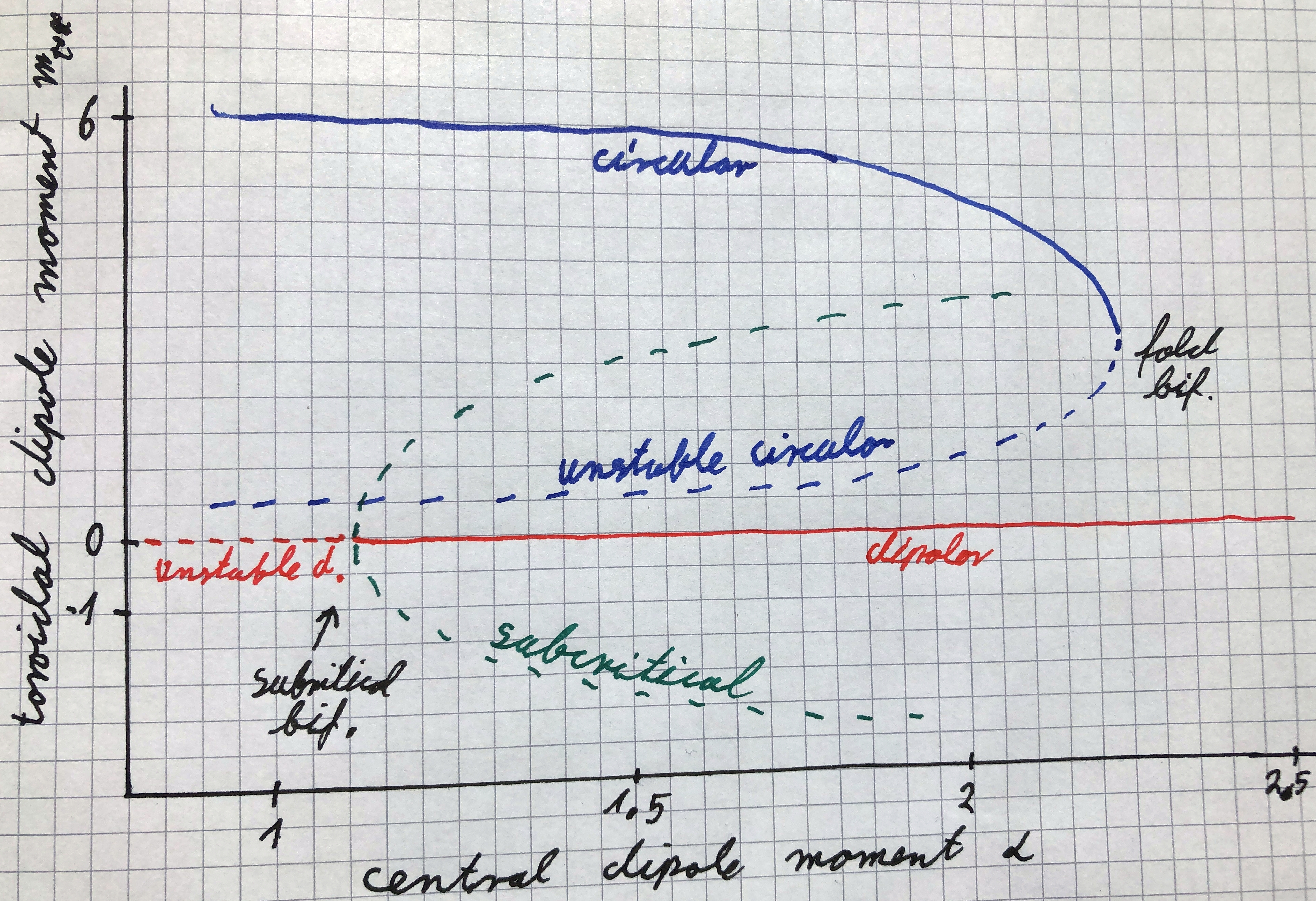}
\captionsetup{labelformat=empty}
\addtocounter{figure}{-1}
\label{graphical abstract}
\end{figure}

\begin{itemize}
    \item Magnetostatic experiment shows almost all fundamental aspects of nonlinear statics.
    \item The toroidal magnetic moment is a perfect order parameter for this magnetic cluster.
    \item Different scaling of fold and subcritical pitchfork bifurcation is unveiled.
    \item Relaxation code allows for interactive exploration of the hysteretic instabilities.
\end{itemize}
\begin{bottompar}

\href{https://doi.org/10.1016/j.jmmm.2022.169520}{https://doi.org/10.1016/j.jmmm.2022.169520}
\end{bottompar}
\end{graphicalabstract}



\begin{keywords}
Magnetostatics\sep
Bifurcation \sep 
Hysteresis \sep
Subcritical bifurcation \sep
Relaxation code \sep
Fold \sep
Cusp catastrophe \sep
Toroidal magnetic moment
\end{keywords}
\maketitle
\section{Introduction}
The natural arrangements of a relatively small number of spherical magnets are a great source of inspiration for experimental classroom demonstrations of magnetostatic problems in general, and bifurcations between static equilibrium states in particular. Smith et al.\ discuss a filled hexagonal arrangement based on seven spherical permanent magnets \cite{Smith2022JMMM}. Their experimental investigation  is compared with a mathematical model based on dipole-dipole interactions, which nicely explains the physics of this magnetostatic problem. Within a certain range of their control parameter $\alpha$ -- the strength of the central dipole relative to the perimeter dipoles of the hexagonal arrangement -- they find bistability. This bistable range is bounded by two instabilities leading to discontinuous jumps of the dipole arrangement. This comment clarifies the nature of these instabilities.
\section{Unveiling the nature of the instabilities}
\begin{figure}
\centering
\includegraphics[width=0.99\linewidth]{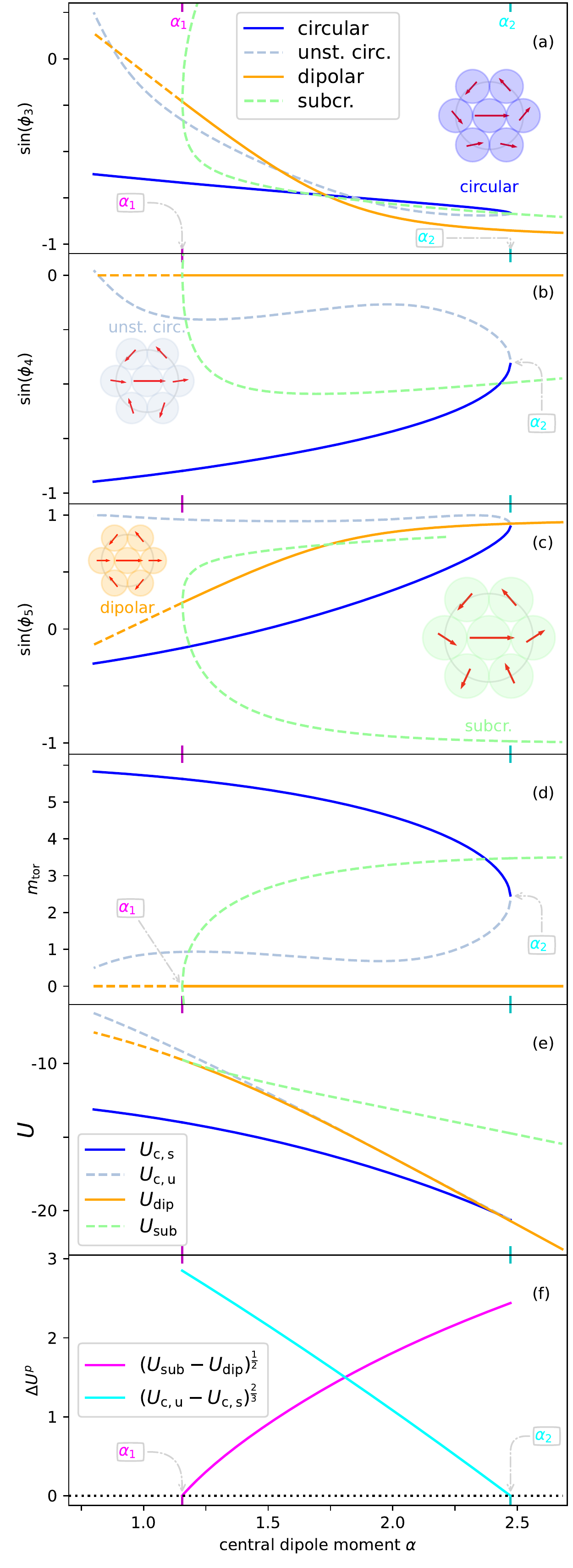}
\caption{The hysteretic jumps between the stable states (solid lines) are explained by showing the unstable branches (dashed).}
\label{Bifurcations}
\end{figure}

In Fig.~\ref{Bifurcations}, we use the nomenclature and color code for the stable branches as introduced in \cite{Smith2022JMMM}. It shows the bistable range of the control parameter $\alpha$ between the two bifurcations at $\alpha_1\approx 1.1545$ and $\alpha_2\approx 2.4724$. The dipolar and the circular state are stable within $\alpha_1<\alpha<\alpha_2$. They were calculated here with a relaxation code that was
developed for studying a similar magnetic cluster problem \cite{Hartung2018}, and is described in detail in another comment \cite{Friedrich2015pre}. Due to the symmetries of these stable states, only three angles, e.\,g., $\phi_{3}$, $\phi_{4}$, and $\phi_{5}$, are necessary to characterize the circular state, and even one angle is sufficient to describe the dipolar state unambiguously. 

In Figs.~\ref{Bifurcations}(a)\,--\,(c), $\phi_{3}$, $\phi_{4}$, and $\phi_{5}$ are shown as a function of $\alpha$. In addition to the two stable branches, we also show two unstable branches that are necessary and sufficient to understand the bifurcations.
The first unstable branch is the continuation of the circular state.
It has been traced using the function 'root' from SciPy \cite{scipy}.
We name it "unstable circular state" and plot the corresponding $\phi(\alpha)$ as a dashed light blue line.
The second unstable state bifurcates at $\alpha_1$ from the dipolar state, breaking its symmetry and destabilizing the highly symmetric dipolar state for $\alpha<\alpha_1$. We name this branch "subcritical branch", because it emerges as a subcritical bifurcation from the dipolar state. The arrangement of all seven dipoles for the four modes of interest within this comment is shown for the middle of the bistable range, namely at $(\alpha_1 +\alpha_2)/2$, and is presented in the form of insets. Both unstable branches have the symmetries of the circular state.

By visual inspection of the $\phi(\alpha)$-curves, the two bifurcations of interest can most clearly be seen in the plot of $\phi_4(\alpha)$. The circular state extends to $\alpha_2$. Of course, it does not simply end here: It folds over into its unstable branch shown as a dashed gray line in this comment. 

The instability of the ensuing dipolar state with $\phi_4=0$ and its higher symmetry ($\phi_5 = -\phi_3$) is also visible here: At $\alpha_1$, a symmetry breaking solution branches off from the dipolar state. It bends over into the stable range of the dipolar solution, i.\,e., it is a subcritical pitchfork bifurcation. Being unstable itself, it destabilizes the dipolar state for $\alpha < \alpha_1$. Such a symmetry breaking subcritical pitchfork bifurcation comes with two branches, which are mirror images of each other with respect to the symmetry they break. In the mirrored branch, $-\phi_4$ replaces $\phi_4$, and $\phi_3$ and $\phi_5$ change their role. Fig.~\ref{Bifurcations} shows only a shorter part of the mirrored branch for clarity.

A natural order parameter for transitions between the di\-po\-lar and the circular state is the toroidal magnetic moment
$\boldsymbol{\mathrm{m}}_{\text{tor}}=\sum_i \boldsymbol{\mathrm{p}}_i \times \boldsymbol{\mathrm{m}}_i$
as described, e.\,g., in Ref.~\cite{Schoenke2015prb}. The z-component of this vector is shown in Fig.~\ref{Bifurcations}(d).
It must start from the value 6 for $\alpha=0$, because the six dipole orientations $\phi_\mathrm{1,..,6}$ are perpendicular to their corresponding location $\boldsymbol{\mathrm{p}}_{1,..,6}$ here. For clarity, the symmetric counterparts with negative toroidal moments are omitted.
The fold can nicely be seen with this order parameter. Fig.~\ref{Bifurcations}(d) also demonstrates that the unstable circular state -- destabilizing the dipolar state at $\alpha_1$ -- does not have the higher symmetry of the dipolar state. The latter can be characterized by $\boldsymbol{\mathrm{m}}_\text{tor}=0$.

The magnetic potential energy $U$ of the four states is shown in Fig.~\ref{Bifurcations}(e). The energies of the dipolar branch $U_\mathrm{dip}$ and the subcritical branch $U_\mathrm{sub}$ meet at $\alpha_1$, and the stable branch $U_\mathrm{c,s}$ of the circular state coincides with the unstable branch $U_\mathrm{c,u}$ at $\alpha_2$. The fundamentally different nature of the two bifurcations leads to distinctive energy difference scalings when approaching the corresponding critical point. For the subcritical bifurcation one expects
    $U_\mathrm{sub} -U_\mathrm{dip} \propto (\alpha - \alpha_1)^2$
by analysis of the corresponding normal form of this bifurcation \cite{strogatz1994nonlinear}. The scaling is best demonstrated visually when plotting the square root of the energy difference in Fig.~\ref{Bifurcations}(f), which leads to a finite slope of the line starting from zero at $\alpha_1$.
For the fold we expect 
    $U_\mathrm{c,u} -U_\mathrm{c,s} \propto (\alpha_2 - \alpha)^{3/2}$,
a semi-cubic parabola, reminiscent of the scaling near a cusp catastrophe \cite{Rehberg1985}. This scaling is best demonstrated visually when plotting the energy difference raised to the power of 2/3, which leads to the corresponding straight line in Fig.~\ref{Bifurcations}(f).
\section{Conclusion}
This experiment is great for classroom demonstrations because it shows many elementary features of static nonlinear systems including spontaneous symmetry breaking, bistability, and subcritical and fold bifurcations with discontinuous jumps. The most precise way to vary $\alpha$ experimentally is presumably the use of an additional coil around the cen\-tral dipole. A pure mechanical modification might be to lower the central dipole from above into the plane of the hexagonal ring. Some implications of the ensuing 3d-sce\-nario are discussed in \cite{Messina_2015}. The bifurcations would change quanti\-ta\-tive\-ly  -- the 3d-aspect has been added to version 1.2.1 of the interactive Python demonstration of this scenario \cite{Rehberg2022}.


\end{document}